\documentclass[aip,jap,notitlepage,reprint]{revtex4-2}

\pdfoutput=1
\usepackage{graphicx}
\usepackage{amssymb}
\usepackage{amsfonts}
\usepackage{amsmath}
\usepackage{hyperref}
\usepackage{mathrsfs}
\usepackage{color}

\newcommand{\bsub}{\begin{subequations}}
\newcommand{\esub}{\end{subequations}}

\newcommand{\sgn}{\operatorname{sgn}}

\newcommand{\beg}{\begin{equation}}
\newcommand{\en}{\end{equation}}

\newcommand \beal  {\begin{eqnarray}\begin{aligned}}
\newcommand \enal  {\end{aligned}\end{eqnarray}}

\newcommand{\m}{\mathbf m}
\newcommand{\n}{\mathbf n}
\newcommand{\h}{\mathbf h}

\newcommand{\abs}[1]{\lvert #1\rvert}

\newcommand{\up}{\uparrow}
\newcommand{\dn}{\downarrow}

\newcommand{\st}{^\ast}

\newcommand{\pmat}{\begin{pmatrix}}
\newcommand{\epmat}{\end{pmatrix}}

\def\w {\omega}

\def\8{\infty}

\def\tt{\frac{2}{3}}

\def\sq2{\sqrt{2}}

\def\d{\partial}

\def\undertext#1{\vtop{\hbox{#1}\kern 1pt \hrule}}
\def\ra{\rightarrow}

\def\pbyp#1#2{\frac{\partial#1}{\partial#2}}

\def\be{\begin{equation}}
\def\ee{\end{equation}}
\def\bea{\begin{eqnarray} & &}
\def\eea{\end{eqnarray}}

\DeclareMathOperator{\sech}{sech}

\begin{document}
\title{Driving a magnetized domain wall in an antiferromagnet by magnons}
\author{Pengtao Shen}
\affiliation{Department of Physics and Astronomy, University of Missouri, Columbia, Missouri 65211, USA}
\author{Yaroslav Tserkovnyak}
\affiliation{Department of Physics and Astronomy, University of California, Los Angeles, California 90095, USA}
\author{Se Kwon Kim}
\affiliation{Department of Physics and Astronomy, University of Missouri, Columbia, Missouri 65211, USA}

\date{\today}

\begin{abstract}
We theoretically study the interaction of magnons, quanta of spin waves, and a domain wall in a one-dimensional easy-axis antiferromagnet in the presence of an external magnetic field applied along the easy axis. To this end, we begin by obtaining the exact solution for spin waves in the background of a domain wall magnetized by an external field. The finite magnetization inside the domain wall is shown to give rise to reflection of magnons scattering off the domain wall, deviating from the well-known result of reflection-free magnons in the absence of a magnetic field. For practical applications of the predicted reflection of magnons, we show that the magnon reflection contributes to the thermally-driven domain-wall motion. Our work leads us to envision that inducing a finite magnetization in antiferromagnetic solitons such as vortices and skyrmions can be used to engender phenomena that do not occur in the absence of magnetization.
\end{abstract} 
\maketitle

\section{introduction}\label{sec:intro}
Magnetic systems can support various topological spin textures such as a domain wall and a vortex, which have been studied for many decades for fundamental interest.~\cite{KosevichPR1990} In particular, the dynamics of a domain wall has been intensively studied for practical applications exemplified by domain-wall racetrack memory~\cite{ParkinScience2008} in the field of spintronics.~\cite{WolfScience2001, ZuticRMP2004} Although conventional material platform for spintronics has been ferromagnets, antiferromagnets have recently emerged as promising material platforms due to their inherent fast dynamics and the absence of the stray field, which can be utilized to realize ultrafast and ultradense spintronic devices.~\cite{JungwirthNat2016,BaltzRMP2018} For this reason, the dynamics of an antiferromagnetic domain wall have been receiving a great attention in the last few years.~\cite{GomonayNP2018} For example, it has been shown that an antiferromagnetic domain wall can be driven by a charge current via spin-transfer torque~\cite{HalsPRL2011, SwavingPRB2011,ChengPRL2014} or via spin-orbit torque.~\cite{GomonayPRL2016, ShiinoPRL2016} Also, a current of magnons, quanta of spin waves, has been shown to be able to induce the dynamics of an antiferromagnetic domain wall by exerting a magnonic force and a magnonic torque.~\cite{TvetenPRL2014, KimPRB2014, SelzerPRL2016, OhPRB2019}

\begin{figure}[!]
	\centering
	\includegraphics[width=0.95\linewidth]{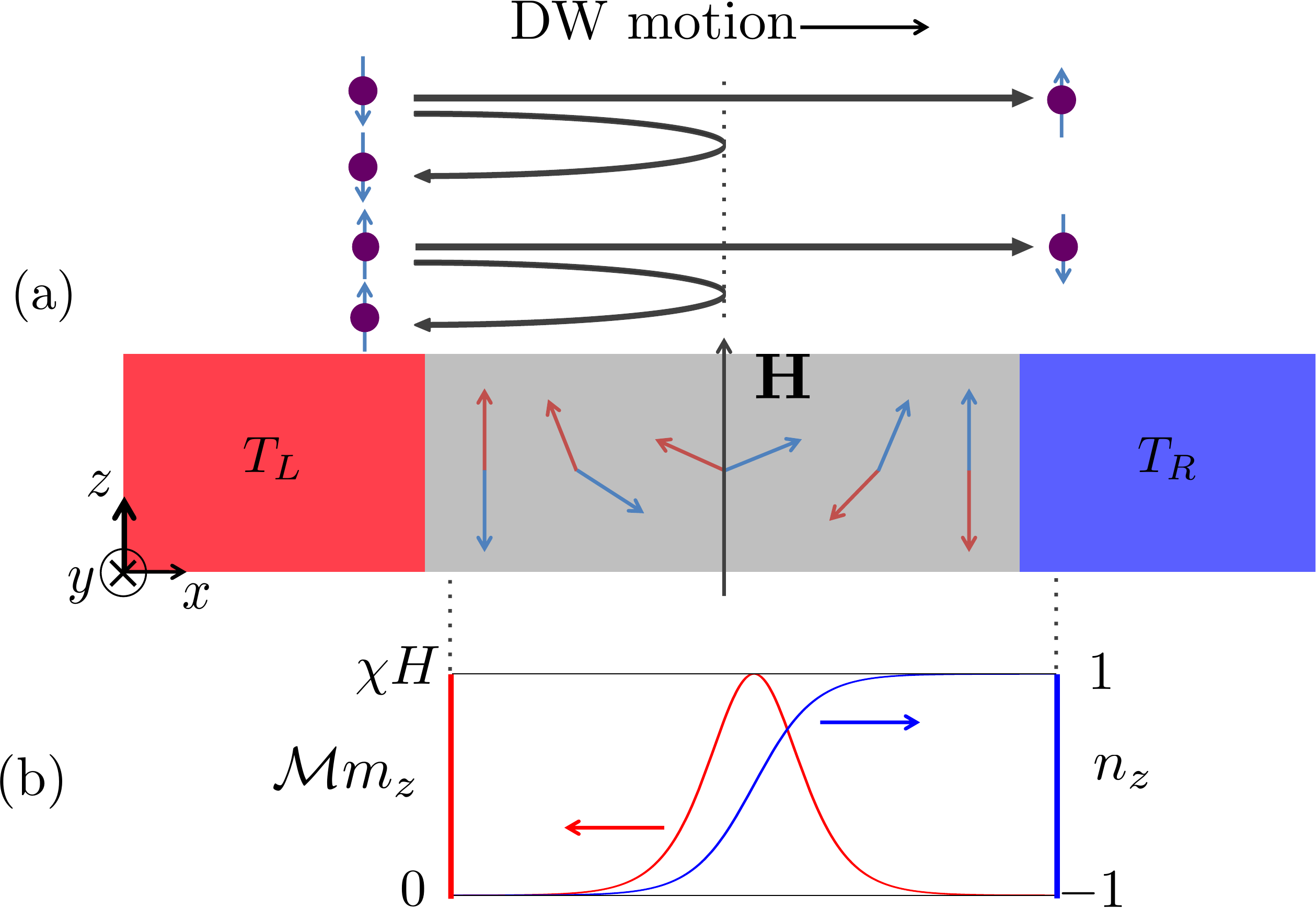}
	\caption{(a) Schematic illustration of the system. A one-dimensional bipartite antiferromagnet with a domain wall is placed between two large thermal reservoirs with constant temperatures $T_L$ and $T_R$ in the presence of an external magnetic field $\mathbf{H} = H \hat{\mathbf{z}}$. The magnetizations of two sublattices are depicted by red and blue arrows. Magnons and their spins are shown as purple dots and blue arrows, respectively. Magnons from the left hotter reservoir incident on the antiferromagnet are partially transmitted and reflected by the magnetized domain wall. A transmitted magnon reverses its spin while a reflected magnon does not. The reflection of magnons transfers their linear momenta to the domain wall and thereby pushes the domain wall to the colder region. (b) The plot shows the $z$ component of staggered magnetization $n_z$ (blue) and uniform magnetization $m_z$ (red). $\mathcal{M}$ and $\chi$ represent the saturation magnetization of one sublattice and the magnetic susceptibility, respectively.}
	\label{fig1}
\end{figure}

The previous researches on the interaction between magnons and antiferromagnetic domain walls have, however, been focused on the situations in which an external magnetic field is absent.~\cite{TvetenPRL2014, KimPRB2014, OhPRB2019} In this paper, we study the interaction of magnons and an antiferromagnetic domain wall in the presence of an external magnetic field, focusing on the effect of the field-induced magnetization of a domain wall on its scattering with magnons. We obtain the exact solutions for a magnetized domain wall and spin waves on top of it within the ``relativistic'' field theory of the non-linear sigma model for antiferromagnets~\cite{HaldanePRL1983, BaryakhtarJETP1983, IvanovPRL1995} by adopting the previous results for a non-magnetized domain wall developed in Ref.~\onlinecite{KimPRB2014}. In particular, as one of the main results of the paper, we find that the field-induced magnetization inside a domain wall engenders reflection of magnons, which can be controlled by varying an external field. We show that the field-induced reflection of magnons contributes to the motion of a domain wall subjected to a thermal bias. The resultant domain-wall velocity is obtained within linear response~\footnote{In this work, we neglect the quantities proportional to $\dot{X}^2$ (domain-wall velocity squared) and $\dot{\Phi}^2$ (angular-velocity squared) by working within the linear response.} using the Landau-B\"{u}ttiker formalism by following the approach taken in Ref.~\onlinecite{YanPRL2012} where a magnon-mediated heat current through a ferromagnetic domain wall is studied. See Fig.~\ref{fig1} for the schematic illustration of the system. 


Our paper is organized as follows. In Sec.~\ref{sec:dwsw}, we present spin-wave solutions on top of a domain wall in a one-dimensional easy-axis antiferromagnet in the presence of an external magnetic field. In Sec.~\ref{sec:dwmotion}, based on the developed theory for the interaction of magnons and a magnetized domain wall, we study the dynamics of a magnetized domain wall driven by a thermal bias. We conclude the paper in Sec.~\ref{sec:disc} by providing summary and discussion.

\section{Domain wall and spin wave}\label{sec:dwsw}
In this section, we obtain an exact solution for a domain wall and a spin wave on top of it in an easy-axis antiferromagnet in the presence of an external magnetic field within the field theory of the non-linear sigma model for antiferromagnets.~\cite{HaldanePRL1983, BaryakhtarJETP1983, IvanovPRL1995}

\subsection{Lagrangian}
We consider an antiferromagnet consisting of two sublattices. The unit magnetizations of the two sublattices are denoted by $\mathbf{m}_1$ and $\mathbf{m}_2$. The Lagrangian density for the antiferromagnet subjected to a uniform magnetic field $\mathbf{H} = H \hat{\mathbf{z}}$ can be written in terms of the staggered magnetization $\n =(\m_1 -\m_2)/2$ and the uniform magnetization $\m = \m_1 + \m_2$ as follows:
\beal\label{lag-nm}
\mathcal L[\n,\m]=&\mathcal{J}\dot\n\cdot(\n\times\m)-\frac{\abs{\mathcal M \m}^2}{2\chi}\\
&+\mathcal M \m\cdot \mathbf H-\mathcal{U}[\n]\,,
\enal
where $\mathcal{J} > 0$ is the density of angular momentum for one sublattice, $ \mathcal M = \gamma\mathcal{J} > 0$ is the magnetization for one sublattice, $\chi > 0$ is the magnetic susceptibility, and $\gamma$ is the gyromagnetic ratio.~\cite{DasguptaPRB2017} Here, the first term is the kinetic term rooted in the spin Berry phase;~\cite{Altland} the second term represents the suppression of the magnetization due to the antiferromagnetic coupling; the third term is the Zeeman coupling; the last term is the potential energy. We also assume that $H > 0$ without loss of generality. We consider an easy-axis antiferromagnet with the potential-energy density given by
\bea
\mathcal U[\n]=\frac{A\abs{\n'}^2-Kn_z^2}2\, ,
\eea
where $A$ is the exchange constant, $'$ represents the spatial derivative with respect to $x$, and $K$ is the strength of easy-axis anisotropy. The uniform magnetization follows the dynamics of staggered magnetization $\m=\chi[\mathcal{J}\dot\n\times\n+\mathcal M\mathbf \n\times(\mathbf{H}\times\n)]/\mathcal M^2$. Correspondingly, the magnetization and the angular momentum density of the antiferromagnet are given by $\mathbf{M} = \mathcal{M} \mathbf{m}$ and $\mathbf{J} = - \mathcal{J} \mathbf{m}$, respectively. As a slave variable of staggered magnetization, the uniform magnetization can be integrated out, which results in the Lagrangian density in terms of only the staggered magnetization: $\mathcal L [\mathbf{n}] =\rho\abs{ \dot \n-\gamma\mathbf H\times\n}^2/2-\mathcal{U}[\n]$, where $\rho=\mathcal J\chi^2$ quantifies the moment of inertia of the staggered magnetization.

It is convenient to use natrual units of length, time and energy,
\bea\label{unit}
\lambda_0=\sqrt{A/K}\,,\quad t_0=\sqrt{\rho/K}\,, \quad \epsilon_0=\sqrt{AK}\, ,
\eea
which will be used hereafter unless otherwise specified. The Lagrangian density is then given by
\bea\label{lag-n2}
\mathcal L=\frac{\abs{ \dot \n}^2-2\dot\n\cdot(\mathbf h\times\n)-\abs{\n'}^2+(1-h^2)n_z^2}2\,,
\eea
where $\h\equiv t_o\gamma \mathbf H=h\hat {\mathbf z}$ with $h > 0$ represents the external magnetic field.

\subsection{Spin waves in a uniform ground state}
The ground states of the antiferromagnet are $\n_0=\sigma \hat {\mathbf z}$,  where $\sigma=\pm1$. To discuss small-amplitude spin-wave excitations on top of a ground state, it is convenient to use a global frame defined by three mutually orthogonal unit vectors: $\hat {\mathbf e}_1$, $\hat {\mathbf e}_2$, $\hat {\mathbf e}_3={\mathbf e}_1\times{\mathbf e}_2=\n_0$. Weakly excited states can be parametrized as $\n (x)=\n_0+\delta\n (x)$ with a small deviation $\delta\n (x)$ orthogonal to $\n_0$. Two fields, $\delta n_1=\delta \n\cdot \hat {\mathbf e}_1$ and $\delta n_2=\delta \n\cdot \hat {\mathbf e}_2$, describe spin waves with linear polarization. We introduce a complex field $\psi=\delta\n\cdot(\hat{\mathbf e}_1+i\hat{\mathbf e}_2)$, which describes spin waves with circular polarization.

Expanding the Lagrangian for small fluctuations in the vicinity of the ground state yields the following spin-wave Lagrangian to the second order in $\psi$:
\bea
\mathcal L_{\text{sw}}=\frac{\abs{\dot\psi}^2-\abs{\psi'}^2-(1-h^2)\abs \psi^2}{2}+ih\psi\st\dot\psi\,.
\eea
The spin-wave equation for a monochromatic wave $\psi(x,t)=\psi(x)\exp(-i\w t)$ is given by
\bea
\w^2\psi=-\psi''+(1-h^2)\psi-2\sigma h\w\,. 
\eea
It has plane-wave solution, $\psi(x)=\Psi \exp(ikx)$, with dispersion relation
\bea
k^2+1-h^2=\w^2+2\sigma h\w\,,
\eea
or equivalently
\bea
k^2+1=(\w+\sigma h)^2\,.
\eea
When $\w<0$, $\delta\n$ precess from $\hat{\mathbf e}_1$ to $\hat{\mathbf e}_2$. We will call such waves right-circularly polarized. The spin waves with $\w>0$ precess from $\hat{\mathbf e}_2$ to $\hat{\mathbf e}_1$ and will be called left-circularly polarized. The spin density and the spin current of the obtained spin-wave solution are give by, respectively,
\bea
\label{eq:spin}
j^0=(-\sigma\w-h)\abs\Psi^2\,,\quad j^1=-\sigma k\abs\Psi^2\,,
\eea
which are reduced to the known results $j^0 = - \sigma \omega |\Psi|^2$ and $j^1 = - \sigma k |\Psi|^2$ when there is no external magnetic field $h = 0$.~\cite{KimPRB2014} The energy density, the energy flux, the linear momentum density, and the pressure for the spin-wave solution are, respectively, given by
\beal
\label{eq:T}
T^{00}&=(\w^2+k^2+(1-h^2))\abs\Psi^2/2=\w(\w+\sigma h)\abs\Psi^2\,,\\
T^{10}&=\w k\abs\Psi^2\,,\\
T^{01}&=(\w k+\sigma h k)\abs\Psi^2\,,\\
T^{11}&=(\w^2+k^2-1+2\sigma h\w)\abs\Psi^2/2=k^2\abs\Psi^2\, .
\enal
The expressions of the spin density, the spin current, and the energy-momentum tensor in terms of the staggered magnetization $\mathbf{n}$, from which Eq.~(\ref{eq:spin}) and Eq.~(\ref{eq:T}) are derived, can be found in Appendix~\ref{apped1}.

\begin{figure}[t]
	\centering
	\includegraphics[width=0.48\linewidth]{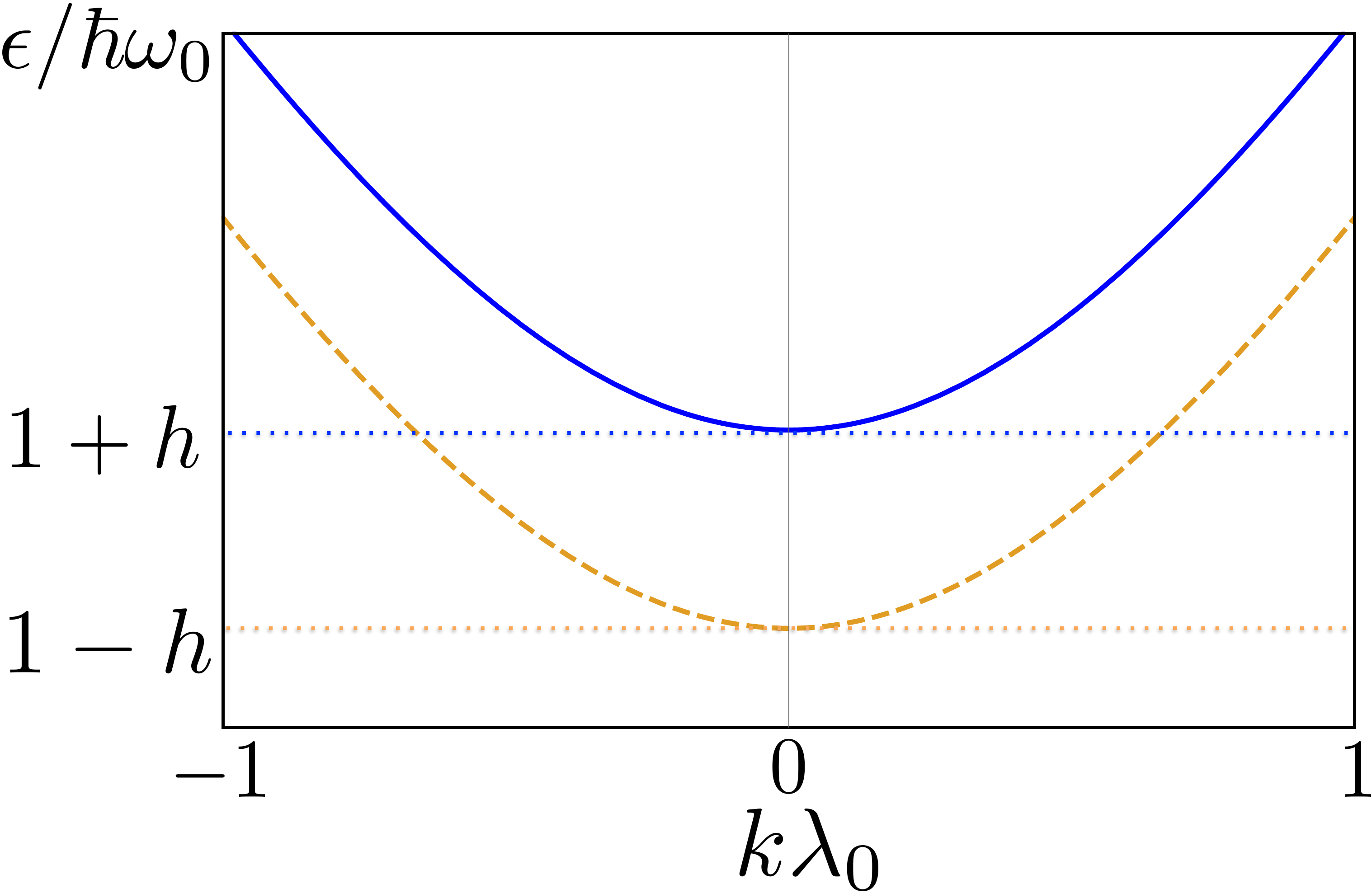}
	\hfill
	\includegraphics[width=0.48\linewidth]{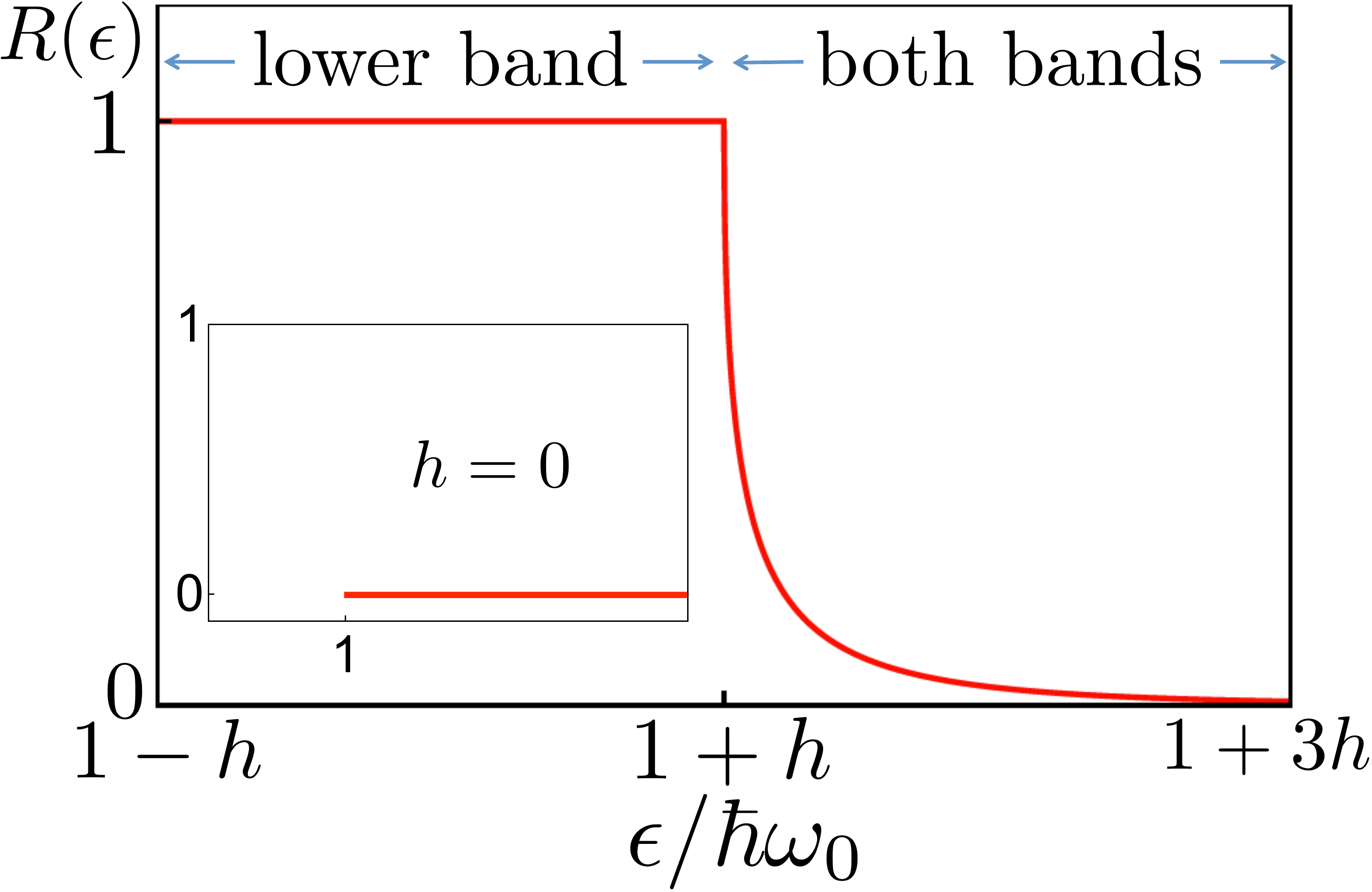}\\
	\qquad\qquad\qquad\, (a)\hfill(b)\qquad\qquad\qquad\quad
	\caption{(a) The dispersion relation of the upper magnon band $\epsilon_+ (k)$ (shown as blue sold line) and the lower magnon band $\epsilon_- (k)$  (shown as yellow dashed line), which have energy gap of $\Delta_\pm=(1\pm h) \hbar \omega_0$, respectively. (b) The dependence of the probability of reflection $R(\epsilon)$ of magnons scattering off a domain wall on the magnon energy $\epsilon$. For energy less than $(1+h) \hbar \omega_0$, magnons exist only in the lower band and they exhibit total reflection: $R(\epsilon)=1$. For energy greater than $(1+h) \hbar \omega_0$,  magnons exist both in the upper and the lower bands and the reflection probability $R(\epsilon)$ is exponentially suppressed as the energy increases. Inset shows the case with no magnetic field $h=0$, when the both bands have the same energy gap and $R(\epsilon)=0$ for any energy. In both plots, $h$ is taken to be 0.1.}
	\label{fig2}
\end{figure}

A quantum of spin waves is referred to as a magnon.~\cite{Kittel} As a boson, a magnon carries angular momentum $\hbar$. Therefore, the magnon number density and current are given by $\abs j^0/\hbar$ and $-\sigma\sgn(\w) j^1/\hbar$. The energy and the linear momentum carried by a single magnon are given by
\beal
\frac{T^{00}}{j^0/\hbar}&=\frac{T^{10}}{-\sigma \sgn(\w) j^1/\hbar}=\hbar \abs \w\,,\\
\frac{T^{01}}{j^0/\hbar}&=\frac{T^{11}}{-\sigma \sgn(\w) j^1/\hbar}=\sgn(\w)\hbar k\,.
\enal
Note that the energy of a magnon is given by the magnitude of the frequency multiplied by the reduced Planck constant: $\epsilon = \hbar |\omega|$. We will omit the reduced Planck constant in the expressions involving the magnon energy when there is no possible confusion. The energies of magnons, whose spin are parallel and anti-parallel to $\mathbf h = h \hat{\mathbf{z}}$ (with $h > 0$ assumed throughout as mentioned above), are respectively given by
\begin{equation}
\epsilon_+(k)=\sqrt{k^2+1}+h \, , \quad \epsilon_-(k)=\sqrt{k^2+1}-h \, .
\end{equation}
Figure~\ref{fig2}(a) shows the dispersion relations of the upper ($+$) and lower ($-$) magnon bands.

\subsection{Domain wall in the presence of a magnetic field}

The exact solution for a static domain wall in the presence of a magnetic field can be obtained by finding a stationary solution of the Lagrangian density [Eq.~(\ref{lag-n2})]:
\beal
\cos\theta(x,t)& =\tanh[\sqrt{1-h^2} (x - X)]\,,\\
\phi(x,t)& =\Phi \, ,
\label{eq:dw}
\enal
where $X$ and $\Phi$ are an arbitrary position and an arbitrary angle of the domain wall. The former and the latter represent zero-energy modes of the domain wall associated with the spontaneous breaking of the translational and the spin-rotational symmetry, respectively.  The external field reduces the domain-wall width by a factor of $\sqrt{1 - h^2}$, which can be considered as a manifestation of the field-induced weakening of effective easy-axis anisotropy. 

To describe the low-energy dynamics of the domain wall, we promote the two zero-energy modes to dynamic variables, $X(t)$ and $\Phi(t)$ in the domain-wall ansatz given in Eq.~(\ref{eq:dw}). The linear momentum of the domain wall can be obtained in terms of the velocity $\dot{X}$ within the collective-coordinate approach~\cite{TretiakovPRL2008, TvetenPRL2013} by integrating the linear-momentum density $T^{10}$ over the system:
\begin{equation}
\label{eq:P}
\begin{split}
P &= \int dx \, T^{10} = - \int dx \, \dot{\mathbf{n}} \cdot \mathbf{n}' \\
&= M_h \dot{X} \, ,
\end{split}
\end{equation}
where $M_h = 2 \sqrt{1 - h^2}$ is the dimensionless mass of the domain wall. In addition, the angular momentum of the domain wall can be obtained in terms of the angular velocity $\dot{\Phi}$ by integrating the spin density $j^0$ over the system:
\begin{equation}
\label{eq:J}
\begin{split}
J &= \int dx \, j^0 = \int dx \, [ \hat {\mathbf z}\cdot(\n\times\dot\n)-h(1 -n_z^2) ] \\
&= I_h (\dot{\Phi} - h) \, ,
\end{split}
\end{equation}
where $I_h = 2 / \sqrt{1 - h^2}$ is the dimensionless moment of inertia of the domain wall. In Sec.~\ref{sec:dwmotion}, we will use the obtained relation [Eq.~(\ref{eq:P})] between the linear momentum and the velocity and the relation [Eq.~(\ref{eq:J})] between the angular momentum and the angular velocity in order to derive the domain-wall velocity driven by magnons. Although we use the above collective-coordinate approach for the dynamics of a domain wall to focus on linear-response regime in the main text, we provide the exact solution for a magnetized domain wall with an arbitrary velocity and an arbitrary angular velocity in Appendix~\ref{apped2}.

\subsection{Spin waves on a static domain wall}
In the presence of a static domain wall $\mathbf{n}_0$ given by Eq.~(\ref{eq:dw}) with $X = 0$, the fluctuation field $\delta\n$ orthogonal to the domain-wall profile can be expanded into a local spin frame: $\hat{\mathbf e}_1=\d\n_0/\d\theta$, $\hat{\mathbf e}_2=\d\n_0/(\sin\theta\d\phi)$, $\hat{\mathbf e}_3=\hat{\mathbf e}_1\times\hat{\mathbf e}_2$.
The Lagrangian density for the complex spin-wave field $\psi=\delta\n\cdot({\mathbf e}_1+i{\mathbf e}_2)$ is given by
\begin{align}
\nonumber \mathcal L_{\text{sw}}=&\frac{\abs{\dot\psi}^2-\abs{\psi'}^2-(1-h^2)[1-2\sech^2 (\sqrt{1-h^2}x)]\abs{\psi}^2}2\\
& +ih\tanh(\sqrt{1-h^2}x)\psi\st\dot\psi\,.
\end{align}
The corresponding spin-wave equation is given by
\beal
0 = & \ddot\psi+2ih\tanh(\sqrt{1-h^2}x)\dot\psi-\psi''\\
&+(1-h^2)[1-2\sech^2 (\sqrt{1-h^2}x)]\psi \, .
\enal
For a monochromatic wave $\psi(x,t)=\psi(x)\exp(-i\w t)$, the wave numbers on the left and the right side of the domain wall ($k_{\text{R/L}}$ for $x\ra\pm\8$ ) are different:
\bea
-\w^2\pm2h\w+k_{\text{R/L}}^2+1-h^2=0\,,
\eea
or
\bea
k_{\text{R/L}}^2+1=(\w\pm h)^2\, ,
\eea
where the $+$ and the $-$ signs correspond to the right (R) and the left (L) sides, respectively. Note that the circular polarization determined by the sign of the frequency $\omega$ is defined in the local frame tied to the staggered magnetization which is reversed by a domain wall. Therefore, the spin direction (with respect to the positive $z$ direction) of magnons depends both on their locations and the sign of the frequency $\omega$: A right-circularly polarized solution ($\w<0$) has $\up$ spin on the left and $\dn$ spin on the right. A left-circularly polarized solution ($\w>0$) has $\dn$ spin on the left and $\up$ spin on the right.

The probability of refection of magnons scattering off the domain wall can be obtained by adopting the known results for magnons on top of a precessing domain wall:~\cite{KimPRB2014, LL3}
\bea
R(\epsilon)=\left\{ 
\begin{array}{ll}
1&\text{if}\,\, \epsilon<\Delta+\\
\frac{\sinh^2[\pi \sqrt{1-h^2}(k_--k_+) / 2]}{\sinh^2[\pi \sqrt{1-h^2}(k_-+k_+) / 2]}&\text{if}\,\,  \epsilon\geq\Delta_+
\end{array}
\,,
\right.
\eea
where $\Delta_\pm = 1\pm h$ is the bottom of the upper $(+)$ and lower$(-)$ branch and $k_\pm^2+1=(\epsilon\mp h)^2$. The probability of transmission is $T(\epsilon) = 1-R(\epsilon)$. This is our first main result: A ``magnetized'' domain wall with $h\neq0$ exhibits reflection of magnons, which does not occur in the absence of a magnetic field.~\cite{IvanovLTP1995, KimPRB2014} Figure~\ref{fig2}(b) shows the plot of the probability of reflection as a function of energy. For energy less than $(1+h) \hbar \omega_0$,  magnons exist only in the lower band and those magnons cannot pass the domain wall since there is no magnon state of the same energy on the other side of the domain wall. Therefore, the magnons whose energy is below $(1 + h) \hbar \omega$ exhibit total reflection: $R(\epsilon)=1$. For energy greater than $(1+h) \hbar \omega_0$, magnons exist both in the upper and the lower bands and thus magnon can pass the domain wall. The reflection probability $R(\epsilon)$ decays exponentially as the energy increases. In the case of $h=0$, the two magnons bands are degenerate and there is no reflection for any energy, $R(\epsilon)=0$, as shown in the previous literarature.~\cite{IvanovLTP1995, KimPRB2014}


\section{motion of a magnetized domain wall driven by a thermal bias}\label{sec:dwmotion}
In this section, we study the motion of a domain wall driven by the reflection of a thermally-induced magnon current. See Fig.~\ref{fig1}(a) for the schematic illustration of the system. We consider the situation where a one-dimensional antiferromagnet chain harboring a domain wall is placed between left and right large thermal reservoirs held at two temperatures $T_L$ and $T_R$, respectively, as shown in Fig.~\ref{fig1}(a), which yields the constant thermal gradient $\partial_x T$. When the relevant inelastic magnon energy-relaxation lengthscale,~\cite{PrakashPRB2018} which we denote by $\lambda_u$, is larger than the domain-wall width, we can consider magnon transport to experience an effective temperature drop of $\delta T \sim \lambda_u \partial_x T$ across the domain wall. Assuming that the system is clean enough so that no elastic magnon momentum scattering occurs on the lengthscale of the domain-wall width, the magnon transport across the domain wall is approximately ballistic. Within this approximation of ballistic magnon transport across the domain wall with the effective temperature drop $\delta T$, we employ the Landau-B\"{u}ttiker formalism to study the thermally-driven domain-wall motion by following the approach taken in Refs.~\onlinecite{MeierPRL2003, YanPRL2012}.

\subsection{Equations of motion for a domain wall}\label{sec:eom}

In the presence of a force $F$ and a torque $\tau$, the equations of motion for a domain-wall position $X$ and the angle $\Phi$ can be derived from Eqs.~(\ref{eq:P}) and (\ref{eq:J}):
\begin{eqnarray}
\dot{P} &=& M_h \ddot X=F - M_h \dot X/t_\text{rel} \, , \\
\dot{J} &=& I_h \ddot\Phi=\tau-I_h \dot\Phi/t_\text{rel} \, ,
\eea
where $t_\text{rel}$ is the phenomenological relaxation time of domain-wall dynamics. Here, the viscous force, $- M_h \dot{X} /t_\text{rel}$, and the viscous torque, $- I_h \dot{\Phi} / t_\text{rel}$, have been added phenomenologically by considering the Rayleigh dissipation function~\cite{Goldstein} $R = \alpha \mathcal{J} \int dx \, \dot{\mathbf{n}}^2$, which yields $R = M_h \dot{X}^2 / (2 t_\text{rel}) + I_h \dot{\Phi}^2 / (2 t_\text{rel})$ with $t_\text{rel} = 1/2 \alpha \mathcal{J}$ when the domain-wall ansatz [Eq.~(\ref{eq:dw})] is plugged in.~\footnote{The Rayleigh dissipation function $R = \alpha \mathcal{J} \int dx \, \dot{\mathbf{n}}^2$ is obtained by simply adding up the Rayleigh dissipation functions of two sublattices, $(\alpha \mathcal{J} /2) \int dx \, \dot{\mathbf{m}}_1^2 + (\alpha \mathcal{J} / 2) \int dx \, \dot{\mathbf{m}}_2^2$, and subsequently extracting the dominant contribution assuming the low-energy dynamics as done in Ref.~\onlinecite{KimPRB2014}.} Here, $\alpha$ is commonly referred to as the Gilbert damping constant characterizing the spin-dissipation rate induced by magnetic dynamics.~\cite{GilbertIEEE2004,TvetenPRL2013,TvetenPRL2014} From the equations of motion, the steady-state solution is given by
\bea\label{stabsol}
\dot{X} = F t_\text{rel} / M_h \,,\quad \dot{\Phi} =\tau t_\text{rel}/I_h\,.
\eea
Below, we will derive the force and the torque induced by magnons driven by a thermal bias.

\subsection{Force exerted by thermal magnons}

Let us first consider a force on a domain wall exerted by $\up$-spin  thermal magnons (upper-energy magnon branch) coming out of the left reservoir, moving to the right. Within the Landauer-B\"{u}ttiker formalism,~\cite{MeierPRL2003,YanPRL2012} the corresponding force is given by 
\begin{equation}
\label{eq:FuL}
F^\up_L = \hbar \int^\8_{1+h} d\epsilon \frac{n_\text B (\beta_\text{L}\epsilon)}{2 \pi} [2R(\epsilon) k_++T(\epsilon)( k_+-k_-)] \, .
\end{equation}
where $n_\text{B}(x) = 1 / (\text e^x - 1)$ is Bose-Einstein distribution function, $\beta_\text{L}=1/(k_\text B T_\text{L})$ is the inverse temperature of the left reservoir, and $k_\pm (\epsilon)=\sqrt{(\epsilon\mp h)^2 - 1}$ is the positive wavevector corresponding to the energy $\epsilon$ for the upper ($+$) and the lower ($-$) magnon branch. On the right-hand side, the first term $2 R(\epsilon) k_+$ represents the force on a domain wall exerted by reflection of $\up$-spin magnons, capturing the transfer of the linear momentum $2 \hbar k_+$ from each reflected magnon to the domain wall. The second term $T(\epsilon) (k_+ - k_-)$ represents the force on a domain wall exerted by $\up$-spin magnons who travel through the domain wall from the left to the right while changing their wavevector from $k_+$ to $k_-$. We would like to mention here that Eq.~(\ref{eq:FuL}) is derived by using $D_\pm(\epsilon) \nu_\pm(\epsilon) =1/2\pi$, which works for one-dimensional systems, where $\nu_\pm(\epsilon) = (1/\hbar) |d \epsilon_\pm / d k|$ is the magnon velocity for the upper ($+$) and the lower ($-$) branch and $D_\pm(\epsilon)=1/[2\pi \nu_\pm(\epsilon)]$ is the magnon density of states.

Analogously, $\dn$-spin magnons (lower-energy magnon branch) from the left reservoir exert the following force on the domain wall:
\begin{equation}
\label{eq:FdL}
F^\dn_L =\hbar \int^\8_{1-h} d\epsilon \frac{n_\text B (\beta_\text{L}\epsilon)}{2 \pi} [2R(\epsilon) k_-+T(\epsilon)( k_--k_+)]\, .
\end{equation}
There are analogous forces $F^\up_R$ and $F^\dn_R$ exerted by $\up$-spin magnons and $\dn$-spin magnons from the right reservoir, which can be obtained from Eq.~(\ref{eq:FuL}) and Eq.~(\ref{eq:FdL}) by replacing $\beta_L$ by $\beta_R$ with the extra factor of $-1$ due to the opposite direction of the force. 

The resultant total force $F = F^\up_L + F^\dn_L + F^\up_R + F^\dn_R$ in physical units instead of natural units [Eq.~(\ref{unit})] is given by
\begin{equation}
\label{Fnum}
\begin{split}
F&=\frac{\hbar\w_0}{\pi\lambda_0}\int^{1+h}_{1-h}d\epsilon~ k_-[n_\text B(\beta_\text{L}\epsilon)-n_\text B(\beta_\text{R}\epsilon)]\\
&+\frac{\hbar\w_0}{\pi\lambda_0}\int^\8_{1+h}d\epsilon~ R(\epsilon) (k_-+k_+)[n_\text B(\beta_\text{L}\epsilon)-n_\text B(\beta_\text{R}\epsilon)]\,.
\end{split}
\end{equation}
The first term in the total force comes from the totally reflected magnons in the lower band whose energies are less than $1 + h$. The second term in the total force comes from the partially reflected magnons in the both upper and lower bands. Note that the contributions from transmitted magnons cancel each other. 

The closed analytical expression for the force can be obtained by assuming sufficiently small temperature difference, $\delta T \ll T_L, T_R$, and sufficiently small magnetic field $|h| \ll 1$. The details of the derivation can be found in Appendix~\ref{app3}. The result is given by 
\bea\label{Fana}
F_{\text{approx.}}=\frac{41}{48} \frac{\hbar\w_0}{\pi\lambda_0}\frac{\hbar\w_0}{T}\frac{\delta T}{T}\frac{(\gamma t_0H)^{3/2}}{\sinh^2(\beta\hbar \w_0)} \, .
\eea
This is our second main result: There is a finite magnonic force on the domain wall in an antiferromagnet subjected to a thermal bias when it is magnetized by an external field.

\subsection{Torque by thermal magnons}
In the left reservoir, there are two types of magnons: the upper branch has $\up$ spin and the lower
branch has $\dn$ spin. A magnon with $\up$ spin and energy $\epsilon$ , which is incident on the domain wall, will transfer the angular momentum 2$\hbar$ to the domain wall after passing it with the probability $T(\epsilon)$. According to the Landauer-B\"{u}ttiker formula,~\cite{MeierPRL2003,YanPRL2012} the torque exerted by $\up$-spin magnons traversing the domain wall is given by
\bea
\tau^\up_L =+2\hbar\int^\8_{1+h}d\epsilon~ n_\text{B}(\beta_\text{L}\epsilon) D_+(\epsilon) \nu_+(\epsilon) T(\epsilon) \,.
\eea
The torque exerted by $\dn$-spin magnons from the left resevoir is given by 
\bea
\tau^\dn_L=-2\hbar\int^\8_{1-h}d\epsilon~n_\text{B}(\beta_\text{L}\epsilon) D_-(\epsilon) \nu_-(\epsilon) T(\epsilon) \,.
\eea
The sum of the two torques is zero:
\bea
\tau=\frac{\hbar\w_0}{\pi}\int^{1+h}_{1-h}d\epsilon~ n_\text B(\beta_\text{L}\epsilon) T(\epsilon) = 0 \, ,
\eea
which can be understood as follows. Only magnons transmitted from the left reservoir to the right reservoir exert the torque on the domain wall. Magnons in the upper branch and the lower branch act the opposite torque with the same magnitude on the domain wall, and, as a result, the torque exerted by magnons coming out of the left reservoir is zero.

There are two more analogous processes involving magnons from the right reservoir, and their sum can be also shown to be zero. Therefore, the total torque on the domain wall is zero. This vanishing of the total torque on the domain wall can be understood by the symmetry argument as explained below.

\begin{figure}[t]
	\centering
	\includegraphics[width=0.7\linewidth]{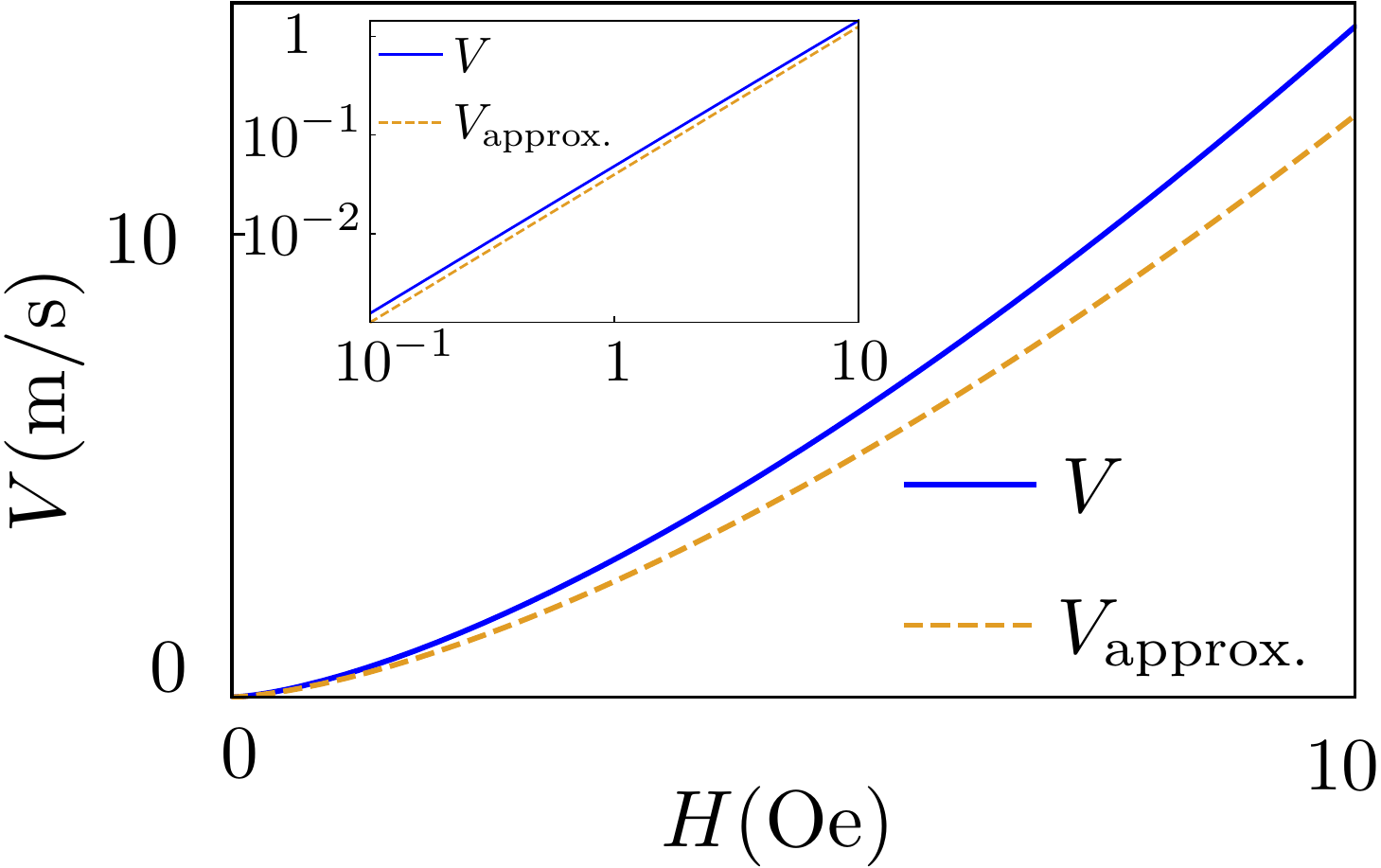}
	\caption{Dependence of a steady-state domain-wall velocity $V$ on the magnetic-field strength $H$.  Parameters used are given in Sec.~\ref{sec:Vsol}. The blue solid line shows the velocity $V = F t_\text{rel} / M_h$ calculated numerically using Eq.~(\ref{Fnum}). The dashed yellow line shows the domain-wall velocity $V = F_\text{approx.} t_\text{rel} / M_h$ calculated analytically using Eq.~(\ref{Fana}). Inset is the log-log plot, which shows $V$ and $V_\text{approx.}$ have the same power dependence on the magnetic field $H$. } 
	\label{fig3}
\end{figure}

\subsection{Steady-state solution}\label{sec:Vsol}
In a steady state, the linear velocity of a domain wall is given by $V = F t_\text{rel} / M_h$ [Eq.~(\ref{stabsol})]. To obtain numerical estimates, we adopt the material parameters used in Ref.~\onlinecite{KimPRB2014}: $\lambda_0$=100~nm, $t_0$=28.4 ps, $t _{\text{rel}}$=25 $t_0$, $\epsilon_0=S\hbar\w_0/2$=1.25 $\hbar\w_0$. For the temperatures, we assume that $T$=100 K and $\delta T=$1 K. The blue solid line in Fig.~\ref{fig3} shows the domain-wall velocity $V$ that is numerically obtained by using Eq.~(\ref{Fnum}) as a force. When we use the  approximate closed expression for the force given in Eq.~(\ref{Fana}), we obtain $V\simeq$ 0.397 m/s$\times$$H^{3/2}$ when the magnetic field $H$ is measured in Oe. This analytical solution for the domain-wall velocity is shown as the dashed yellow line in Fig.~\ref{fig3}. One can see that the numerical result [Eq.~(\ref{Fnum})] and the analytical result [Eq.~(\ref{Fana})] agree well for small magnetic fields.

The steady-state angular velocity of a domain wall is zero due to the vanishing torque. This can be understood by the symmetry argument assuming that the antiferromagnet respects the structural inversion symmetry. The magnetic field $\mathbf{H}$ and the angular velocity of the domain wall $\dot{\Phi}$ (which is given by $\hat{\mathbf{z}} \cdot (\mathbf{n} \times \dot{\mathbf{n}})$ at the domain-wall center) are even under the inversion operator, but the temperature difference $\delta T \propto \partial_x T$ is odd under the inversion since the positions of the hotter region and the colder region are switched. The even-parity quantity $\dot{\Phi}$ cannot linearly depend on the odd-parity quantity $\delta T$ in the inversion-symmetric antiferromagnet, and thus that it should vanish to linear order in $\delta T$, which agrees with the previous explicit derivation based on magnonic torque.

\section{summary and discussion}\label{sec:disc}
We have studied the interaction of magnons and a domain wall in a one-dimensional antiferromagnet in the presence of an external field within the field theory. We have shown that a magnon can be partially reflected by a magnetized domain wall even when the domain wall is static. We have utilized the obtained reflection of magnons incident on a domain wall to predict the motion of a magnetized domain wall when it is subjected to a thermal bias. 

In the presence of a temperature gradient, it is known that an antiferromagnetic domain wall can move also by the Brownian motion.~\cite{KimPRB2015Thermophoresis, YanPRB2018} The corresponding Brownian force on the domain wall can be approximated by $F_\text{B} \sim k_B \partial_x T$.~\cite{KimPRB2015Thermophoresis} For sufficiently high temperatures $k_B T \gg \hbar \omega_0$, the ratio of the magnon-induced force obtained in this work to the Brownian force can be estimated as $F_\text{approx.} / F_\text{B} \sim (\lambda_u / \lambda_0) (\gamma t_0 H)^{3/2} (\sigma / d^2)$, where $\lambda_u$ is the lengthscale of magnon-energy relaxation (appearing in the effective temperature drop experienced by magnons traveling across the domain wall, $\delta T \sim \lambda_u \partial_x T$), $\sigma$ is the crosssection of the antiferromagnet wire, $d$ is the lattice constant of the antiferromagnet, and thus $\sigma / d^2$ represents the number of magnon modes per unit length. The magnon-induced force is expected to dominate the Brownian force either when the crosssection of the antiferromagnet wire is sufficiently large or when the applied magnetic field is sufficiently strong. For example, when $\sigma = 100$ nm$^2$ and $d = 0.5$ nm are used for structural parameters, $t_0 = 28.4$ ps is used for the characteristic time scale for the antiferromagnet as in Sec.~\ref{sec:Vsol}, and $\lambda_u = 300$ nm is used for the magnon energy-relaxation length (adopted from the result for yttrium-iron-garnet reported in Ref.~\onlinecite{PrakashPRB2018}), then the magnon-induced force is expected to dominate the Brownian force for magnetic fields $H \gg 20$ Oe. In addition, since magnon reflections are found to be appreciable only for low-energy magnons, high-energy thermal magnons whose wavelength is shorter than the domain-wall width do not contribute to our main results significantly. However, they may become relevant when some magnon-relaxation processes (beyond the ballistic transport assumed in this work) become important in the context of the conventional magnonic spin torques acting on smooth magnetic textures~\cite{YanPRL2011} such as the entropic torque studied in Ref.~\onlinecite{SchlickeiserPRL2014, KimPRB2015}.

In this paper, we have focused on the effect of the field-induced magnetization on the interaction of an antiferromagnetic domain wall and magnons. However, a domain wall is just one member of a large class of topological solitons that exist in antiferromagnets. We therefore envision that applying an external magnetic field to other antiferromagnetic solitons such as skyrmions~\cite{SkyrmePRSA1961, IvanovJETP1986, 	GomonayNP2018} and vortices may give rise to phenomena that do not occur for non-magnetized solitons.

\begin{acknowledgments}
This work is supported by the University of Missouri (P.S. and S.K.K.) and by the U.S. Department of Energy, Office of Basic Energy Sciences under Award No. {DE-SC0012190} (Y.T.). S.K.K. acknowledges Young Investigator Grant (YIG) from Korean-American Scientists and Engineers Association (KSEA). 
\end{acknowledgments}

\appendix
\section{Energy-momentum tensor, spin density, and spin current}\label{apped1}
In this appendix, we discuss several physical quantities of interest, which can be obtained within the classical field theory.~\cite{Goldstein, KimPRB2014} For the given staggered magnetization $\n$, the energy density, the energy flux, the linear momentum density, and the pressure are given by, respectively,
\beal
T^{00}&=\dot\n\cdot\pbyp{\mathcal{L}}{\dot\n}-\mathcal L=\frac{\abs{ \dot \n}^2+\abs{\n'}^2-(1-h^2)n_z^2}2\,,\\
T^{10}&=\dot\n\cdot\pbyp{\mathcal{L}}{\n'}=-\dot\n\cdot\n'\,,\\
T^{01}&=-\n'\cdot\pbyp{\mathcal{L}}{\dot\n}=-\dot\n\cdot\n'+\mathbf h\cdot(\n\times\n')\,,\\
T^{11}&=-\n'\cdot\pbyp{\mathcal{L}}{\n'}+\mathcal L\\
&=\frac{\abs{ \dot \n}^2-2\dot\n\cdot(\mathbf h\times\n)+\abs{\n'}^2+(1-h^2)n_z^2}2\,.
\enal
The spin density and the spin current are respectively given by
\beal
j^0=&\pbyp{\mathcal{L}}{\dot\n}\cdot(\hat {\mathbf z}\times\n)=\hat {\mathbf z}\cdot(\n\times\dot\n)- h(1-n_z^2) \,,\\
j^1=&\pbyp{\mathcal{L}}{\n'}\cdot(\hat {\mathbf z}\times\n)=-\hat {\mathbf z}\cdot (\n\times \n')\,.
\enal

\section{Domain-wall solution in the presence of a magnetic field}\label{apped2}
In this appendix, we derive a domain-wall solution in the presence of magnetic field. The unit vector field $\mathbf{n}$ can be written in terms of angle variables $\theta$ and $\phi$, $\n(x,t)=(\sin\theta\cos\phi,\sin\theta\sin\phi,\cos\theta)$ and $\d_i\n=\d_i\theta\hat\theta+\sin\theta\d_i\phi\hat\phi$, where $\hat\theta={\d\n}/{\d\theta}=(\cos\theta\cos\phi,\cos\theta\sin\phi,-\sin\theta)$ and $\hat\phi={\d\n}/(\d\sin\theta\d\phi)=(-\sin\phi,\cos\phi,0)$.
The Lagrangian density Eq.~(\ref{lag-n2}) is then given by
\beal\label{lag-ang}
\mathcal L=\frac{\dot\theta^2+\sin^2\theta(\dot\phi-h)^2-\theta'^2-\sin^2\theta\phi'^2-\sin^2\theta}2\,.
\enal
The domain-wall solution in the absence of a magnetic field $h = 0$ is well known.~\cite{MikeskaJPC1980, BaryakhtarJETP1983, HaldanePRL1983} The exact solution for a domain wall in the presence of the external field $h \neq 0$ can be obtained from the aforementioned solution with $h = 0$ by the following transformation:
\bea 
\label{eq:tr}
\theta(x,t)\to\theta_h(x,t)\,,\quad
\phi(x,t)\to\phi_h(x,t)-ht\, ,
\eea
and it is given in Eq.~(\ref{eq:dw}).

Next, we discuss an exact domain-wall solution with a finite velocity and a finite angular velocity in the presence of an external field. When the external field is absent $h = 0$, the exact solution for a domain wall for boundary condition $\n(\pm\8)=\pm\hat {\mathbf z}$ which is moving
at linear velocity $V$ and rotating at angular velocity $\Omega$ in its rest frame is given by~\cite{HaldanePRL1983, IvanovPRL1995,KimPRB2014}
\beal
&\cos\theta(x,t)=\tanh \left[ \frac{\sqrt{1-\Omega^2}(x-Vt)}{\sqrt{1-V^2}} \right] \,,\\
&\phi(x,t)=\Omega\frac{t-Vx}{\sqrt{(1-V^2)}} \,.
\enal
Here, we would like to mention that the angular velocity of a domain wall in the lab frame is given not by $\Omega$, but by $\dot{\phi} = \Omega / \sqrt{1 - V^2}$. Therefore, $\Omega$ should be considered as a parameter characterizing the angular velocity, not as the angular velocity itself. The energy, the momentum and the angular momentum of the domain wall are respectively given by
\beal
E&=\int T^{00}dx=\frac{M_0}{\sqrt{1-V^2}\sqrt{1-\Omega^2}}\,,\\
P&=\int T^{10}dx=\frac{M_0 V}{\sqrt{1-V^2}\sqrt{1-\Omega^2}}\,,\\
J&=\int j^0dx=\frac{I_0 \Omega}{\sqrt{1-\Omega^2}}\,,
\enal
where the mass $M_0=2$ and the moment of inertia $I_0=2$.

The exact solution for a domain wall in the presence of an external field $h \neq 0$ can be obtained through the aforementioned transformation [Eq.~(\ref{eq:tr})]:
\beal
&\cos\theta(x,t)=\tanh\left[ \frac{\sqrt{1-\Omega^2}(x-Vt)}{\sqrt{1-V^2}} \right]\,,\\
&\phi(x,t)=\Omega\frac{t-Vx}{\sqrt{(1-V^2)}}+ht\,.
\enal
Note that the angular velocity of the domain wall is given by $\dot{\phi} = h + \Omega / \sqrt{1 - V^2}$. The energy, the momentum and the angular momentum of the domain wall are respectively given by
\beal
E&=\frac{M_0}{\sqrt{1-V^2}\sqrt{1-\Omega^2}}+\frac{M_0 h\Omega}{\sqrt{1-\Omega^2}}\,,\\
P&=\frac{M_0 V}{\sqrt{1-V^2}\sqrt{1-\Omega^2}}+\frac{M_0 h\Omega V}{\sqrt{1-\Omega^2}}\,,\\
J&=\frac{I_0 \Omega}{\sqrt{1-\Omega^2}}\,.
\enal

The result for the linear momentum within the collective-coordinate approach, $P = M_h V$ [Eq.~(\ref{eq:P})] with $M_h = M_0 \sqrt{1 -h^2}$, can be obtained from the above expression for $P$ by replacing $\Omega$ by $-h$ and neglecting $V^2$. In addition, the result for the angular momentum, $J = I_h (\dot{\Phi}-h)$ [Eq.~(\ref{eq:J})] with $I_h = I_0/\sqrt{1 -h^2}$ , can be obtained from the above expression for $J$ by replacing $\Omega^2$ in the denominator by $h^2$.

\section{Analytical expression for the force}
\label{app3}

Assuming that the temperature difference is sufficiently small, $\delta T \ll T_\text L, T_\text R$, we can simplify the expression for the total force [Eq.~(\ref{Fnum})] by using 
$n_\text B(\beta_\text{L}\epsilon)-n_\text B(\beta_\text{R}\epsilon)\simeq - (\delta T / T^2) n'_B(\beta\epsilon)\epsilon$, where $T=(T_\text L +T_\text R)/2$ is the average temperature. For small $|h| \ll 1$, the first term in the force [Eq.~(\ref{Fnum})] can be approximated by
\bea
F_1\simeq\frac{\hbar\w_0}{\pi\lambda_0}\frac{\hbar\w_0}{T}\frac{\delta T}{T}\frac{1}{\sinh^2(\beta\hbar \w_0)}\frac 2 3h^{3/2}\,.
\eea
The $h^{3/2}$ power dependence is a result of the multiplicative effect of the gap difference $\sim h$ and the average linear momentum transferred $\sim h^{1/2}$. In the limit of a vanishing magnetic field $h\to0$, the gap difference between upper band and lower band disappears as shown in Fig.~\ref{fig2}(b), and, thus $F_1$ contribution vanishes.

The second term in Eq.~(\ref{Fnum}) involves $R(\epsilon)$ which is exponentially suppressed as the magnon energy $\epsilon$ is far above the upper gap $\Delta_+ = 1 + h$: $R(\epsilon) \approx \exp[-\sqrt{8(\epsilon-1-h)/h}]$. Let us write $\epsilon=1+h+\delta \epsilon$ with $\delta \epsilon > 0$. Then, the dominant contribution to $F_2$ comes form the magnons with long wavelengths, $\delta\epsilon\ll h$. In this approximation, we can set $k_1\simeq \sqrt{4 h}$, $k_+\simeq\sqrt{2\delta\epsilon}$, and $\epsilon\simeq1$ in the integrand. This results in the following approximation for $F_2$:
\bea
F_2\simeq \frac{\hbar\w_0}{\pi\lambda_0}\frac{\hbar\w_0}{T}\frac{\delta T}{T}\frac{3}{16\sinh^2(\beta\hbar \w_0)}h^{3/2}\,.
\eea
The $h^{3/2}$ power dependence is a result of the multiplicative effect of the exponential decay length $\sim h$ and average momentum transferred $\sim h^{1/2}$. In the limit of a vanishing magnetic field $h\to0$, $R(\epsilon)=0$ for all energy, and thus $F_2$ contribution also vanishes. The sum of the above two analytical expressions yield Eq.~(\ref{Fana}) in the main text.

\bibliography{bibfile}

\end{document}